\documentclass{dhbenelux}

\usepackage{booktabs} 
\usepackage{authblk} 
\usepackage{graphicx} 
\usepackage{amsmath}
\usepackage{adjustbox}

\usepackage[style=apa, backend=biber]{biblatex}
\DeclareLanguageMapping{english}{english-apa}
\author[1,2]{Guoyang Rong}
\author[2,3]{Ying Chen\thanks{Corresponding author: Ying Chen (matcheny@nus.edu.sg)}}
\author[4]{Thorsten Koch}
\author[5]{Keisuke Honda}

\affil[1]{School of Information Management, Wuhan University, 430072, China}
\affil[2]{Center for Quantitative Finance, Department of Mathematics, National University of Singapore, 119076, Singapore}
\affil[3]{Risk Management Institute, National University of Singapore, 119076, Singapore}
\affil[4]{Zuse Institute Berlin \& Technische Universit\"at Berlin, 14195, Germany}
\affil[5]{The Institute of Statistical Mathematics, 190-8562, Japan}

\title{From Coverage to Prestige: A Comprehensive Assessment of Large-Scale Scientometric Data}

\begin{document}

\maketitle

\begin{abstract}
    \noindent {As research in the Scientometric deepens, the impact of data quality on research outcomes has garnered increasing attention. This study, based on Web of Science (WoS) and Crossref datasets, systematically evaluates the differences between data sources and the effects of data merging through matching, comparison, and integration. Two core metrics were employed: Reference Coverage Rate (RCR) and Article Scientific Prestige (ASP), which respectively measure citation completeness (quantity) and academic influence (quality). The results indicate that the WoS dataset outperforms Crossref in its coverage of high-impact literature and ASP scores, while the Crossref dataset provides complementary value through its broader coverage of literature. Data merging significantly improves the completeness of the citation network, with particularly pronounced benefits in smaller disciplinary clusters such as Education and Arts. However, data merging also introduces some low-quality citations, resulting in a polarization of overall data quality. Moreover, the impact of data merging varies across disciplines; high-impact clusters such as Science, Biology, and Medicine benefit the most, whereas clusters like Social Sciences and Arts are more vulnerable to negative effects. This study highlights the critical role of data sources in Scientometric research and provides a framework for assessing and improving data quality.}
\end{abstract}


\section{INTRODUCTION}

The Science of Science and Scientometric has attracted growing research interest due to the increasing availability of large-scale datasets that enable researchers to explore how science operates and where innovation occurs \parencite{Lin2023,Liu2023,Zeng2017,Azoulay2018,clauset2017}. The analysis of scientific citations heavily relies on datasets, with different datasets often leading to varying conclusions. The quality of a dataset encompasses factors such as the completeness of journals, papers, information, citation relationships, and disciplinary classifications. Any issues with data quality may affect computational results and researchers' interpretations of conclusions and discussion. Studying the quality of large-scale datasets not only assists researchers in selecting suitable datasets for future studies but also enables peer reviewers to evaluate the credibility of research findings. 

\textcite{Lin2023} summarized 30 data sources commonly used in the Science of Science research and published SciSciNet processed dataset, based on the original data from Microsoft Academic Graph (MAG), while \textcite{Wang2020} introduced MAG and discussed how it can be utilized in analytics. \textcite{Baas2020} analyzed the advantages of Scopus as a data source for quantitative scientific research. \textcite{Birkle2020} presented methodologies and cases for using Web of Science (WoS) as a data source for studying scientific and academic activities. \textcite{Herzog2020} introduced the Dimensions, detailing its inclusion of data types such as funding, publications, citations, clinical trials, patents, and policy documents, as well as its potential role in scientometrics research. \textcite{Hendricks2020} provided an overview of the composition, management, and updates of academic metadata collected and provided by Crossref. \textcite{Peroni2020} introduced the tools, services, and datasets provided by OpenCitations for scientometric research. These studies provide detailed descriptions of the types, characteristics, and access of the datasets but rarely address the quality of them and lack comparative analyses between different datasets. This study aims to address the following research questions to fill this gap. \vspace{\baselineskip}

\noindent \textbf{RQ1.} How significant are the differences between citation datasets from different sources?

\noindent \textbf{RQ2.} Can merging large-scale datasets improve the quality of the data?

\noindent \textbf{RQ3.} What are the impacts of merging data on different disciplinary clusters? \vspace{\baselineskip}

We selected Web of Science (WoS) and Crossref as our datasets because they complement each other in creating a balanced and comprehensive bibliometric dataset, which can  illustrate the methods for assessing data completeness and quality in this study. WoS is known as a benchmark for quality, with rigorous selection criteria that focus on high-impact, peer-reviewed journals in established disciplines like natural sciences, social sciences, and engineering. Its curated metadata and emphasis on citation prestige make it ideal for analyzing traditional academic impact. In contrast, CrossRef offers breadth and diversity, capturing content from niche journals, and emerging interdisciplinary fields often overlooked by WoS. It includes data from smaller and regional publishers, ensuring broader geographical and linguistic representation. While WoS excels in depth and quality, CrossRef provides inclusivity and coverage of non-traditional research areas. 

To address the research questions, we first compared Crossref and WoS to identify the differences between their datasets, then merged the two datasets to assess whether the integration improved data completeness and quality. Finally, we analyzed the impact of the merged dataset on Science of Science research from cluster perspectives.

\section{METHODS}

\subsection{Data Matching and Merging}

To compare differences in the same papers included in different datasets, we first matched the papers in the two datasets using their DOIs. Crossref uses DOI as the UID to index article records, but the references are not obliged to include DOI. WoS includes DOI as a data attribute, 39\% of the records in our dataset lack DOI. Additionally, 4\% of the DOIs in WoS could not be matched with Crossref. For papers that could not be matched using DOIs, we performed matching based on publication titles and journal ISSNs.

\begin{table}[h!]
\centering
\caption{Statistical summary of datasets}
\label{tab_data_summary}
\small
\begin{tabular}{p{3cm}p{1cm}ccc}
\toprule
\textbf{} & \textbf{} & \textbf{Web of Science} & \textbf{Crossref} & \textbf{Merged} \\
\midrule
Temporal Span   &    & 1981--2020       & 1901--2022   &  1901--2022  \\
Papers          &    & 63,092,811       & 94,334,766   &  119,020,655 \\
Non-Ref. Papers &    & 12,364,404       & 47,683,386   &  56,898,385  \\
Cites           & Max.   & 29,906         & 41,712       &  46,546      \\
                & Avg.   & 8.29           & 7.79         &  8.19        \\
                & Median & 5.00           & 3.00         &  3.00        \\
References      & Max.   & 6,481          & 4,805        &  6,481       \\
                & Avg.   & 22.12          & 24.65        &  23.40       \\
                & Median & 20.00          & 18.00        &  16.00       \\
\bottomrule
\end{tabular}
\end{table}

This method matched 38,422,068 papers between the two datasets, representing 60.9\% of the WoS and 40.72\% of the Crossref dataset. We assigned new unique identifiers to the matched papers and proceeded to reference Matching. We processed the references of both the matched and original datasets in parallel. Shared references between the two datasets were matched using the aforementioned method, while other references were linked to the matched papers to establish new citation relationships. The detailed methods and processes for data matching and merging are outlined in our previous research \parencite{yueksel2024}. We integrated the unmatched data from WOS and Crossref with the matched data to create a large-scale dataset. Table \ref{tab_data_summary} presents statistics of Papers, Cites, and References of the WoS, Crossref, and merged data. 

\subsection{Dataset Comparison Methods}

We employed two approaches to evaluate the quantity and quality of the datasets to address RQ1. First, the quantity of citations serves as the straightforward indicator of the dataset's coverage of citation relationships. Due to the coverage limitations of databases, the references included in any database are unlikely to fully align with the actual citations. For the same paper in different datasets, a higher number of recorded citations indicates more comprehensive data coverage. Therefore, we used the reference coverage rate (RCR) to evaluate the comprehensiveness of the datasets in capturing citation relationships.

\begin{equation}
\text{RCR}_i^{\text{ds}} =  \frac{\text{R}_i^{\text{ds}}}{\text{R}_i^{\text{m}}}
\label{eq:RCR}
\end{equation}
Where $\text{R}_i^{\text{ds}}$ represents the number of references for article $i$ in dataset $ds$, and $\text{R}_i^{\text{m}}$ represents the number of references for article $i$ in the merged dataset, their ratio reflects the completeness of references for article $i$ in dataset $ds$. The closer the value of $\text{R}_i^{\text{m}}$ is to the actual number of references cited by a paper, the more accurate the RCR becomes. Since obtaining the actual reference counts is challenging, we use the number of references in the merged dataset as a closer approximation to the true values. An RCR value closer to 1 indicates that the citation data included in the dataset is more comprehensive.

Second, we used the Article's Scientific Prestige (ASP) metric from our previous research to measure the impact of each article \parencite{chen2023}. Based on eigenvector centrality, ASP accounts for direct and indirect citations. Its effectiveness was validated using WoS data in multidisciplinary large-scale citation networks. The ASP of an article $i$ in a citation network of size $N$ is defined as follows:

\begin{equation}
\text{ASP}_i = (1 - d) + d \sum_{j=1}^{N} \text{ASP}_j \frac{L_{ij}}{m_j}
\label{eq:ASP}
\end{equation} where $L_{ij} = 1$ if an article $j$ cites an article $i$ and $L_{ij} = 0$ otherwise, $m_j = \sum_k L_{kj}$ is the total number of articles that $j$ links to. In other words, the ratio $L_{ij} / m_j$ denotes the fraction of references article $j$ has cited. The damping factor $d$ influences how much “prestige” of an article is passed on to the references. If $d$ is larger, more is passed on to the referenced (older) articles. As $d$ gets smaller, the benefit of being cited decreases. The minimum value of ASP is $1 - d$, which means that the article is not cited. For the same paper, a higher ASP across different datasets reflects the inclusion of high-impact papers and their associated citations.

RCR and ASP provide two perspectives for evaluating dataset quantity and quality to address RQ1, focusing on coverage comprehensiveness and inclusion of high-quality papers.

\subsection{Evaluation of Merging Performance}

To address RQ2, we evaluated changes before and after merging by analyzing the variations in rankings based on citations and ASP. The changes in citations reflect the comprehensiveness of the citation coverage in the database before and after integration. ASP evaluates the quality of multi-level citations, and it can indicate the quality of citation. We used the difference in rankings between them to quantify the quality of newly added citations in the data set after integration. The ranking changes can be categorized into four cases, summarized by the \textbf{GOLD} framework: 
\begin{itemize}
    \item[-]\textbf{G}lorious Upgrade, where merging introduces high-quality data, enhancing both citation and ASP rankings; 
    \item[-] \textbf{O}rdinary Gain, reflecting improved citation completeness but relatively low prestige of the added citations; 
    \item[-] \textbf{L}ow Impact, indicating a decline in overall quality due to the introduction of low-impact citations; and 
    \item[-] \textbf{D}iverse Boost, which represents a mixed outcome with some high-quality additions despite a generally low baseline citation quality.
\end{itemize}
Specifically, the case Glorious Upgrade represents the most desirable outcome, where the merging process significantly enhances the dataset with high-quality data. Articles in the original database demonstrate high citation quality, indicated by being referenced by a larger number of influential papers. After merging, the data reveals that the difference between the ASP ranking (measuring prestige through multi-level citations) and the citation ranking widens, suggesting the addition of high-quality new citations or improvement in the quality of existing ones. Furthermore, if the ASP ranking improves post-merge, it provides additional evidence of enhanced data quality. Quantitatively, this is expressed as:
\[
\text{\textbf{G}lorious Upgrade:} \quad \big(n_\text{ds} < k_\text{ds} \quad  \text{and} \quad (k_\text{m} - n_\text{m}) - (k_\text{ds} - n_\text{ds}) \geq 0\big) \quad \text{or} \quad n_\text{m} \leq n_\text{ds}
\label{eq:c1}
\]
where \( n_\text{ds} \) and \( n_\text{m} \) denote the ASP rankings of a single article in the original dataset and the merged dataset, respectively. \( k_\text{ds} \) and \( k_\text{m} \) represent the citation rankings of the same article in the original and merged datasets, respectively.  The condition \( n_\text{ds} < k_\text{ds} \) indicates that the ASP ranking is higher (closer to the top) than the citation ranking in the original dataset. The term \( (k_\text{m} - n_\text{m}) - (k_\text{ds} - n_\text{ds}) \geq 0 \) quantifies the widening gap between ASP and citation rankings after merging, reflecting the improvement in data quality. Additionally, \( n_\text{m} \leq n_\text{ds} \) suggests that the ASP ranking improves after merging, further underscoring the enhanced quality of the integrated dataset.

The case Ordinary Gain represents a scenario where the merging process improves citation completeness but does not significantly enhance the scientific prestige of the dataset. 
\begin{equation}
\text{\textbf{O}rdinary Gain:} \quad n_\text{ds} < k_\text{ds} \quad \text{and} \quad (k_\text{m} - n_\text{m}) - (k_\text{ds} - n_\text{ds}) < 0 \quad \text{and} \quad n_\text{m} > n_\text{ds}
\label{eq:c2}
\end{equation} 
In contrast to the case Glorious Upgrade, the reduced difference between the higher original ASP ranking and the citation ranking suggests that, although data integration improved the completeness of the citation, the scientific prestige of the newly added citations is relatively low. Additionally, the ASP ranking worsens after merging, further reflecting the limited impact of the new data on enhancing overall quality.

\begin{equation}
\text{\textbf{L}ow Impact:} \quad n_\text{ds} \geq k_\text{ds} \quad \text{and} \quad (k_\text{m} - n_\text{m}) - (k_\text{ds} - n_\text{ds}) < 0 \quad \text{and} \quad n_\text{m} > n_\text{ds}
\label{eq:c4}
\end{equation} \(n_\text{ds} \geq k_\text{ds}\) indicates that the quality of citations for this article in the original dataset is relatively low, as the citation ranking is better compared to the ASP ranking. Furthermore, \( (k_\text{m} - n_\text{m}) - (k_\text{ds} - n_\text{ds}) < 0 \) suggests that after merging, the gap between ASP rankings and citation rankings has widened. This indicates that the integration introduced more of low-impact citations, reflecting a further decline in the overall quality of the dataset.

\begin{equation}
\text{\textbf{D}iverse Boost:} \quad n_\text{ds} \geq k_\text{ds} \quad \text{and} \quad (k_\text{m} - n_\text{m}) - (k_\text{ds} - n_\text{ds}) \geq 0 \quad \text{and} \quad n_\text{m} > n_\text{ds}
\label{eq:c3}
\end{equation} 

In contrast to the Case Low Impact, although the overall quality of this article's citations is relatively low, the integration includes new high-quality citations. Diverse Boost represents a mixed outcome, characterized by some high-quality additions despite a generally low baseline citation quality.

\section{RESULTS}

\subsection{Dataset Comparison}

To address RQ1, we compare the quality of the datasets from two perspectives: completeness (quantity) and article prestige (quality). Evaluating these aspects is crucial because the reliability of scientometric research depends heavily on the accuracy and comprehensiveness of the underlying data. A thorough comparison ensures that the datasets can support valid conclusions and provide a solid foundation for further analyses.

\subsubsection{Dataset Completeness}

We used the RCR metric to evaluate the completeness of the citations included in the two datasets. The comparison is restricted to the 38,422,068 papers that matched. The summary statistics of RCR are presented in Table \ref{tab_RCR}. 

\begin{table}[h!]
\centering
\caption{Statistical summary of RCR}
\label{tab_RCR}
\small
\begin{tabular}{p{1cm}p{1cm}ll}
\toprule
\textbf{} & \textbf{} & \textbf{Web of Science} & \textbf{Crossref}  \\
\midrule
$\text{RCR}_i^{\text{ds}}$ &  Range  & [0 \hspace{0.5cm} 1.000]       & [0 \hspace{0.5cm} 1.000]     \\
                           &  Avg.   & 0.837           & 0.748         \\
                           &  Median & 0.917           & 0.907         \\
\bottomrule
\end{tabular}
\end{table}

In Table \ref{tab_RCR}, the average RCR for Web of Science is 0.837, which is significantly higher than the 0.748 observed for Crossref. This indicates that Web of Science demonstrates a higher level of overall citation coverage, enabling a more comprehensive recording of citation relationships among publications. Additionally, the median RCR for Web of Science is 0.917, compared to 0.907 for Crossref. Although the difference is relatively small, it still suggests that for over half of the publications, the Web of Science dataset offers slightly better stability in citation coverage than Crossref.

\begin{figure}[h!]
    \centering
    \includegraphics[width=\textwidth]{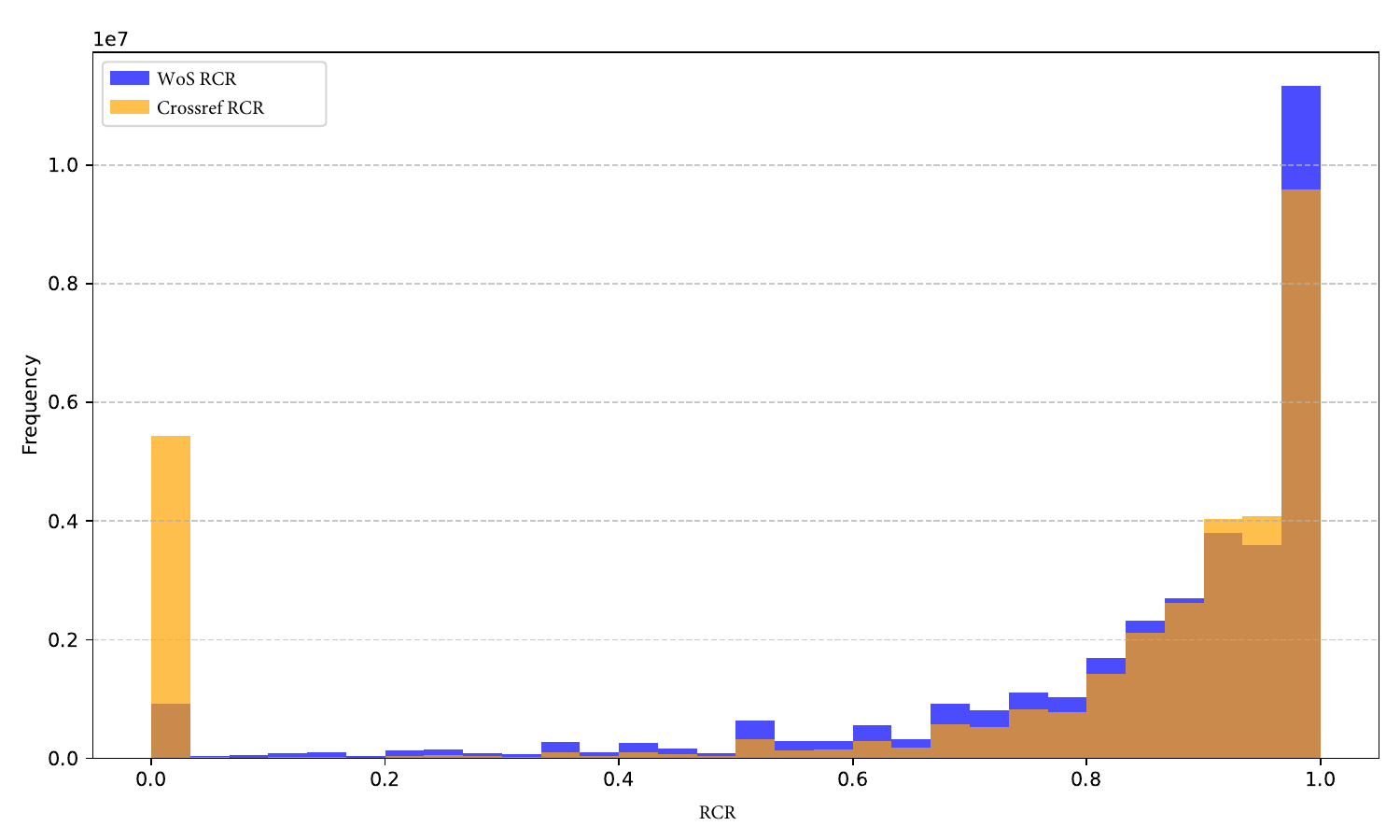}
    \caption{Distribution of RCR for Wos and CrossRef}
    \label{Fig_RCR}
\end{figure}

Figure \ref{Fig_RCR} illustrates the distribution of RCR across the two datasets. An RCR value of 1 indicates that the dataset has fully covered the citations within our sample, reflecting the inclusion of complete reference data. In this regard, WoS outperforms Crossref. Conversely, an RCR value of 0 indicates that the dataset has not included any references for the article, demonstrating poor completeness. This issue is more prevalent in Crossref than in WoS. It is worth noting that, as articles without references in both datasets were excluded prior to the calculation of RCR, an RCR of 0 in Figure \ref{Fig_RCR} indicates that the references for that article are included in the other dataset. Furthermore, Crossref shows a higher distribution of articles in the [0.9, 1) range compared to WoS, indicating that Crossref demonstrates a certain advantage when it comes to nearly complete citation coverage.

\subsubsection{Article Prestige}

We use the ASP metric to evaluate the impact of article in datasets. The calculation of ASP includes all articles with references from both datasets, comprising 50,728,407 articles from WoS and 46,651,380 articles from Crossref. Based on the preliminary research and testing \parencite{chen2023}, we selected a five-year time window for calculating the ASP. The summary statistics of ASP are presented in Table \ref{tab_ASP}.

\begin{table}[h!]
\centering
\caption{Statistical summary of ASP}
\label{tab_ASP}
\small
\begin{tabular}{p{1cm}p{1cm}ll}
\toprule
\textbf{} & \textbf{} & \textbf{Web of Science} & \textbf{Crossref}  \\
\midrule
ASP                        &  Range  & [0.50 \hspace{0.5cm} 2,746.82]  & [0.50 \hspace{0.5cm} 1,930.12 ] \\
                           &  Avg.   & 0.83           & 0.72           \\
                           &  Median & 0.54           & 0.50           \\
\bottomrule
\end{tabular}
\end{table}

In Table \ref{tab_ASP}, the range of ASP for WoS has a higher maximum value compared to Crossref, indicating that the most influential papers in our study are found in the WoS dataset. The average ASP for WoS is 0.83, slightly higher than Crossref's 0.72, suggesting that the WoS dataset contains articles with higher average impact. Although the median ASP values for the two datasets differ only slightly, Crossref's median is 0.50, indicating that at least half of the articles in the Crossref dataset have not been cited. These data indicate that WoS generally outperforms Crossref in terms of the impact of the articles it includes. To further analyze the inclusion of high-impact articles in both datasets.

\begin{table}[h!]
\caption{The sets of top 20 articles ranked by ASP from WoS or Crossref}
\label{tab_ASP_rank}
\small
\begin{tabular}{cccccp{7cm}}
\toprule
\textbf{ASP\textsubscript{WoS}} & \textbf{n\textsubscript{WoS}} & \textbf{ASP\textsubscript{Crossref}} & \textbf{n\textsubscript{Crossref}} & \textbf{Year} & \textbf{Title} \\
\midrule
2,746.82 & 1   & 1,930.12 & 1     & 2008 & A Short History of SHELX... \\
1,202.68 & 2   & 881.33   & 6     & 2013 & MEGA6: Molecular Evolutionary Genetics... \\
1,161.44 & 3   & None      & None  & 2016 & Deep Residual Learning for Image... \\
781.19   & 4   & 1,140.10 & 3     & 2016 & MEGA7: Molecular Evolutionary Genetics... \\
654.87   & 5   & 571.49   & 17    & 2007 & MEGA4: Molecular Evolutionary Genetics... \\
623.80   & 6   & 653.30   & 10    & 1986 & Possible High-Tc Superconductivity... \\
621.57   & 7   & 277.78   & 49    & 2011 & LIBSVM: A Library for Support Vector... \\
584.07   & 8   & None      & None  & 2015 & Going Deeper with... \\
583.11   & 9   & 65.57    & 807   & 2003 & Single-Crystal Structure Validation... \\
564.27   & 10  & 720.21   & 8     & 2015 & Deep Learning... \\
562.11   & 11  & 646.00   & 12    & 2015 & Fitting Linear Mixed-Effects Models Using... \\
550.26   & 12  & 0.55     & 36,418,443 & 2011 & Global Cancer Statistics... \\
529.51   & 13  & 267.40   & 52    & 2004 & MEGA3: Integrated Software for Molecular... \\
493.05   & 14  & 242.29   & 59    & 2017 & ImageNet Classification with Deep... \\
480.46   & 15  & 628.08   & 13    & 2015 & Crystal Structure Refinement with... \\
463.84   & 16  & 421.98   & 24    & 1987 & Superconductivity at 93-K in a... \\
436.78   & 17  & 458.50   & 22    & 2015 & Global Cancer Statistics,... \\
429.20   & 18  & 90.14    & 425   & 1988 & Primer-Directed Enzymatic Amplification... \\
395.23   & 19  & None      & None  & 2012 & Convincing Evidence from Controlled and... \\
393.35   & 20  & 425.16   & 23    & 2009 & Structure Validation in Chemical... \\
358.38   & 24  & 1,069.74 & 5     & 2018 & Global Cancer Statistics 2018: GLOBOCAN... \\
351.13   & 26  & 1,380.61 & 2     & 2020 & Clinical Features of Patients Infected with... \\
290.99   & 40  & 531.71   & 18    & 2011 & Global Cancer Statistics... \\
262.11   & 47  & 867.17   & 7     & 2020 & A Novel Coronavirus from Patients with... \\
207.64   & 77  & 516.44   & 19    & 2020 & Clinical Characteristics of 138... \\
175.01   & 105 & 646.41   & 11    & 2020 & Early Transmission Dynamics in Wuhan... \\
168.52   & 116 & 601.88   & 14    & 2020 & Epidemiological and Clinical Characteristics... \\
168.52   & 117 & 601.49   & 15    & 2020 & Clinical Characteristics of Coronavirus... \\
158.34   & 139 & 589.73   & 16    & 2018 & MEGA X: Molecular Evolutionary Genetics... \\
142.11   & 178 & 657.72   & 9     & 2020 & A Familial Cluster of Pneumonia Associated... \\
121.02   & 268 & 511.01   & 20    & 2020 & Clinical Course and Risk Factors for... \\
11.74    & 42,029 & 1,131.98 & 4  & 2011 & MEGA5: Molecular Evolutionary Genetics... \\
\midrule
\multicolumn{6}{p{14cm}}{\textit{Note: "None" indicates that the article is not indexed in the corresponding dataset.}} \\
\bottomrule
\end{tabular}
\end{table}

Table \ref{tab_ASP_rank} presents the top 20 articles ranked by ASP from either the Web of Science (WoS) or Crossref datasets, with n\textsubscript{WoS} representing the rankings in the WoS dataset and n\textsubscript{Crossref} indicating the rankings in the Crossref dataset. The results reveal that, despite employing the same evaluation method, substantial discrepancies can occur depending on the dataset utilized.

Firstly, certain high-quality papers indexed in WoS are not included in Crossref. For example, the article “Deep Residual Learning for Image Recognition” has a high ASP of 1,161.44  and ranks third (n\textsubscript{WoS} = 3) in WoS, but it is not recorded in Crossref. Similarly, “Going Deeper with Convolutions” ranks 8th in WoS with an ASP value of 584.07 but is also absent in Crossref. These examples highlight significant differences in the coverage of high-quality papers between databases.

Secondly, even when the same paper is included in both datasets, its ranking can vary substantially. For instance, “Global Cancer Statistics (vol 61, pg 69, 2011)” ranks 12th in WoS (n\textsubscript{WoS} = 12) but is ranked far lower in Crossref (n\textsubscript{Crossref} = 36,418,443). Conversely, “Clinical Characteristics of 138 Hospitalized Patients With 2019 Novel Coronavirus–Infected Pneumonia in Wuhan, China” ranks 77th in WoS but rises to 19th in Crossref. Such discrepancies result in divergent academic evaluations for the same paper depending on the database used.

Finally, even when a paper holds the same ranking in both datasets, its ASP values can differ significantly. For example, “A Short History of SHELX” is ranked 1st in both WoS and Crossref. However, its ASP value in WoS is 2,746.82, whereas in Crossref it is 1,930.12. These differences underscore the impact of data quality inconsistencies across databases, which can significantly affect the evaluation of academic research outcomes. 

These discrepancies highlight significant differences between WoS and Crossref in terms of data coverage and the assessment of paper impact. Relying solely on either database may result in incomplete conclusions. Therefore, investigating the impact of merging large-scale datasets on data quality is necessary.

\subsection{Data Merge Performance}

We analyzed the performance of the merged data from four perspectives. First, we analyze the statistical data of various disciplines in the merged dataset. Second, we compare the changes in citation counts, ASP, and cluster calculation results before and after data merging. Third, we evaluate the effectiveness of data merging in terms of data completeness. Finally, we assess the impact of data merging on data quality.

\subsubsection{Merged Data Statistics}

We adopted the method reported in our previous study \parencite{chen2023} to group WoS classifications into 14 clusters. Crossref does not provide disciplinary classifications for papers. Therefore, we employed a three-step process to match papers from the Crossref dataset into the 14 clusters. First, we matched the overlapping papers from the Crossref dataset with the WoS dataset and directly applied the existing cluster results. This process assigned 38,422,068 matched papers to the clusters. Second, we created a word list to classify not matched journals included in the Crossref dataset, which comprises 97,288 journals. For these journals, we extracted keywords from their titles and mapped them to corresponding clusters. For instance, the word 'mathematics' was categorized under 'Science', and 'surgery' was categorized under 'Medicine'. Using this word list, we successfully matched 64,068 journals. Third, we assigned 67,731,612 papers to clusters based on the journal list, leaving 12,866,975 papers unclassified. We further analyzed the unmatched papers and found that their average citation count was 1.32, significantly lower than the merged dataset's average of 8.19. This suggests that the influence of the unmatched papers is relatively low and can be disregarded in our analysis.

\begin{table}[h!]
\centering
\caption{Merged statistical summary of Cit and ASP at cluster level}
\label{tab_summary_cluster}
\small
\begin{tabular}{llllllll}
\toprule
\textbf{Cluster} & \multicolumn{3}{c}{\textbf{\#Cites}} & \multicolumn{3}{c}{\textbf{ASP}} \\
\cmidrule(lr){2-4} \cmidrule(lr){5-7}
                 & \textbf{Range}  & \textbf{Median} & \textbf{Mean}  & \textbf{Range}  & \textbf{Median} & \textbf{Mean}  \\
\midrule
Medicine         & [0, 46,546]     & 0               & 5.37           & [0.5, 1,405.52] & 0.50             & 0.75           \\
Science          & [0, 32,296]     & 2               & 7.61           & [0.5, 3,054.65] & 0.59            & 0.85           \\
Biology          & [0, 29,781]     & 2               & 7.29           & [0.5, 1,631.57] & 0.56            & 0.79           \\
Engineering      & [0, 3,224]      & 0               & 4.18           & [0.5, 287.57]   & 0.50             & 0.73           \\
Social Science   & [0, 12,406]     & 0               & 1.69           & [0.5, 829.06]   & 0.50            & 0.60           \\
Geography        & [0, 4,859]      & 1               & 5.21           & [0.5, 204.16]   & 0.54            & 0.75           \\
Arts             & [0, 1,430]      & 0               & 0.43           & [0.5, 100.88]   & 0.50            & 0.54           \\
Computer Science & [0, 24,075]     & 0               & 3.13           & [0.5, 1,358.85] & 0.50             & 0.73           \\
Psychology       & [0, 8,523]      & 0               & 4.26           & [0.5, 427.89]   & 0.50             & 0.76           \\
Management       & [0, 3,636]      & 0               & 3.23           & [0.5, 186.16]   & 0.50             & 0.72           \\
Law and Policy   & [0, 3,288]      & 0               & 1.10           & [0.5, 154.64]   & 0.50             & 0.57           \\
Building         & [0, 970]        & 0               & 3.24           & [0.5, 36.23]    & 0.50             & 0.69           \\
Education        & [0, 2,314]      & 0               & 1.45           & [0.5, 141.90]   & 0.50             & 0.60           \\
City Development & [0, 1,345]      & 0               & 1.57           & [0.5, 79.83]    & 0.50             & 0.59           \\
\bottomrule
\end{tabular}
\end{table}

\begin{figure}[h!]
    \centering
    \includegraphics[width=\textwidth]{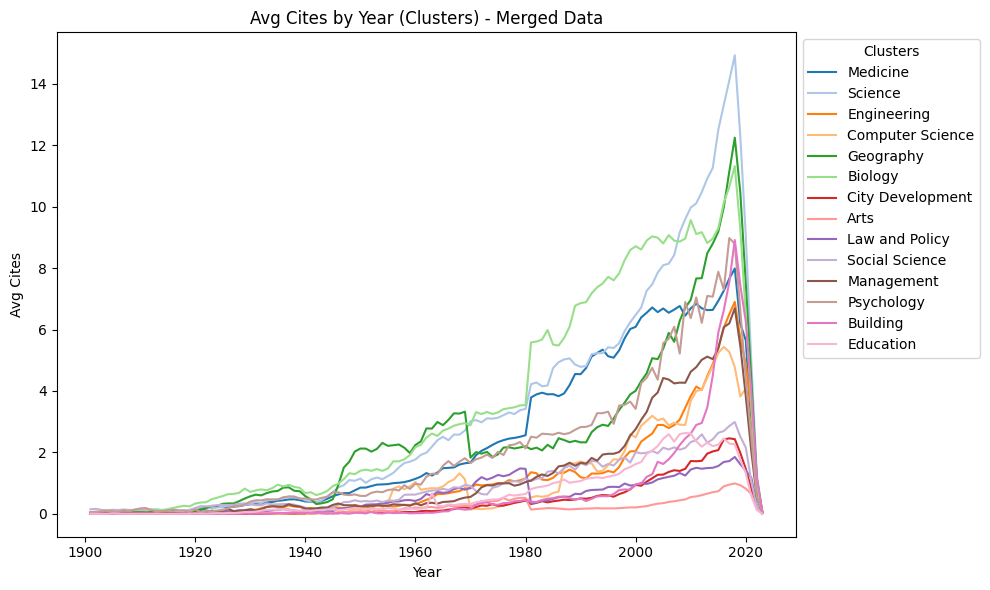}
    \caption{Annual average citation counts in cluster level}
    \label{Fig_Merged_Cluster_Cit}
\end{figure}

\begin{figure}[h!]
    \centering
    \includegraphics[width=\textwidth]{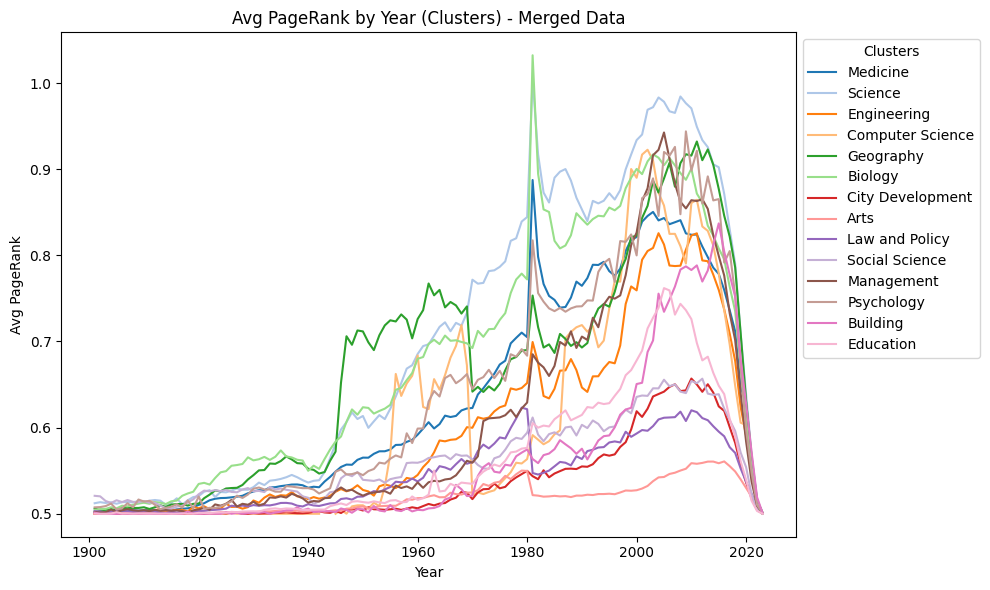}
    \caption{Annual average ASP in cluster level}
    \label{Fig_Merged_Cluster_ASP}
\end{figure}

Table \ref{tab_summary_cluster} provides a statistical summary of citation counts and ASP metrics across different disciplinary clusters. In terms of citation counts, the Biology has the highest average, reaching 13.80, with a median of 7. Medicine and Science follow closely, with average citation counts of 11.70 and 11.00, respectively, reflecting their notable advantage in citation numbers. Fields such as Arts and Building have relatively low average citation counts, at 2.40 and 4.00, respectively, highlighting the challenges these disciplines face in academic dissemination. For the ASP metric, Biology and Science also stand out with high averages of 0.96 and 0.95, respectively, suggesting strong academic influence in these fields. In contrast, disciplines such as Engineering (0.79), Social Sciences (0.60), and Arts (0.53) show lower ASP averages, indicating relatively weaker overall academic influence and citation activity in these areas. In comparison, although Medicine exhibits relatively high citation counts, its ASP metric is not as strong as its citation performance suggests. This indicates that, compared to other clusters at a similar level, the quality of citations in the Medicine field is relatively lower. In contrast, Geography's ASP is relatively higher than its citation count, indicating that although Geography lacks an advantage in citation quantity, the quality of its citations is high. Additionally, Figure \ref{Fig_Merged_Cluster_Cit} and Figure \ref{Fig_Merged_Cluster_ASP} illustrate the annual average citation counts and average ASP values for these clusters, respectively.

\subsubsection{Performance Before and After Merging}

\begin{figure}[h!]
    \centering
    \includegraphics[width=\textwidth]{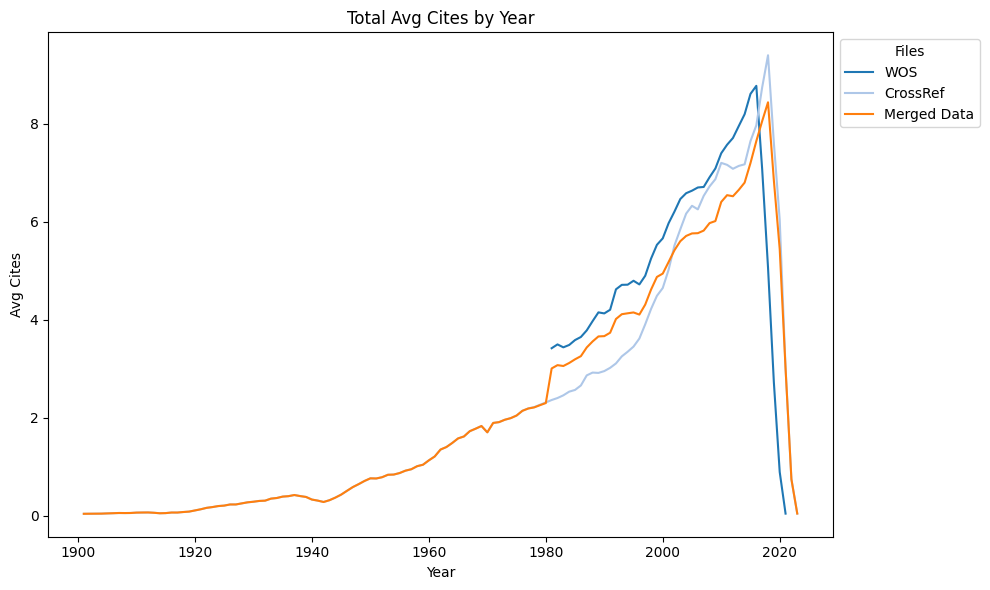}
    \caption{Annual average citation counts}
    \label{Fig_Merged_Avg_Cites}
\end{figure}

Figure \ref{Fig_Merged_Avg_Cites} presents a comparison of the annual average citation counts across the WoS, Crossref, and merged data. The two original datasets cover different time periods. Before 1981, only data from Crossref is available; therefore, the merged dataset from 1901 to 1980 overlaps entirely with the Crossref data. In 1981, influenced by the inclusion of the WoS dataset, the completeness of the merged data improved significantly. As a result, the average citation count showed a notable increase compared to the Crossref data. In 2000, the annual average citation count of the merged data began to fall below that of the two original datasets. This phenomenon can be attributed to two main factors: (1) Around the year 2000, there was a sharp increase in the volume of published papers. Both the WoS and Crossref datasets exhibited a pronounced long-tail effect in citation distributions, characterized by a vast majority of low-citation or zero-citation papers dominating the datasets, with only a small fraction of highly cited papers included. (2) The two datasets contained significant overlaps in highly cited papers. Consequently, after merging, the total number of highly cited papers showed only a modest increase. In contrast, the datasets included a large number of unique low-citation and zero-citation papers. This exacerbated the long-tail effect in the merged data, further diminishing the average citation count.

In addition, the average citation count in both the WoS and Crossref datasets shows a sharp decline approaching 2020, which inevitably affects the merged data. This is because it takes time for papers to accumulate citations, and our dataset only includes publications up to 2020. Consequently, papers published close to 2020 do not had sufficient time to garner a substantial number of citations.

\begin{figure}[h!]
    \centering
    \includegraphics[width=\textwidth]{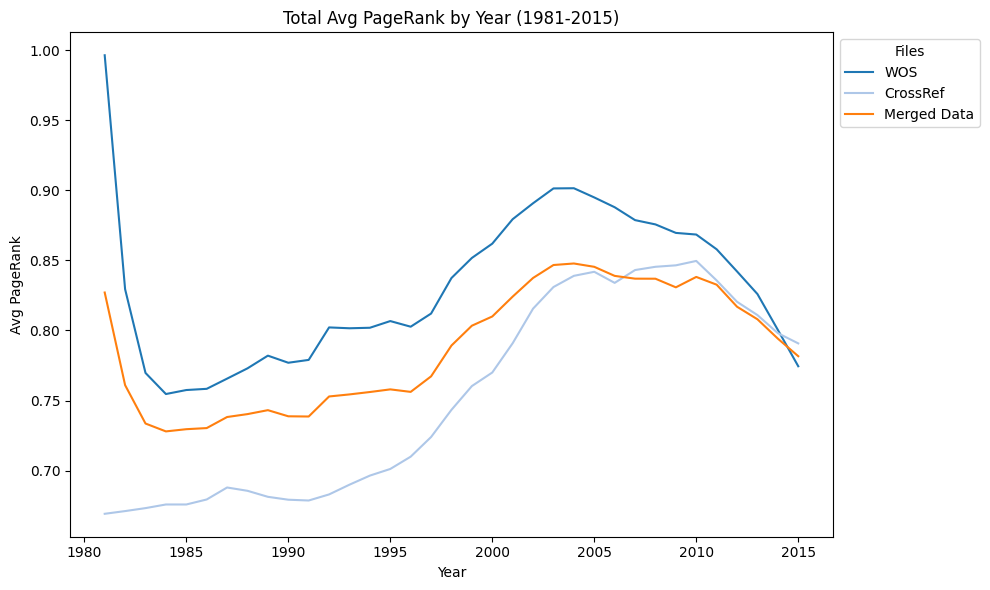}
    \caption{Annual average ASP}
    \label{Fig_Merged_Avg_ASP}
\end{figure}

Figure \ref{Fig_Merged_Avg_ASP} presents a comparison of the annual average ASP across the WoS, Crossref, and merged data. We selected the period from 1981 to 2015 for analysis for two main reasons. First, as shown in Figure \ref{Fig_Merged_Avg_Cites}, the merged data from before 1981 and after 2020 overlaps with the Crossref data, making it redundant for inclusion. Second, since we use a five-year time window for calculating the ASP, and our dataset extends only to 2020, years after 2015 lack sufficient data to complete the five-year window. As a result, the ASP calculated for these years would not adhere to a consistent standard compared to earlier years. Additionally, due to the influence of the time window, the ASP of papers published near 2015 accounts for citations from papers up to 2020. As a result, the ASP of papers near 2015 shows a declining trend earlier than their citation counts.

The results of ASP calculation differ from citation count in two key aspects. First, in 1981, the average ASP exhibited an abnormal surge. This can be attributed to the characteristics of the WoS dataset during its initial year of inclusion. At this time, the WoS dataset had a relatively small data volume, an unstable network structure, and highly concentrated citations. Since the PageRank algorithm typically requires iteration within a large and well-connected network to converge, the smaller number of nodes in the early network led to anomalous ASP values. As the network size expanded over time, the ASP gradually stabilized. Second, the ASP of the merged data is influenced by the long-tail effect later than the citation count. While the citation count began to exhibit the impact of data merging and the resulting long-tail effect around the year 2000, this influence only became apparent for the ASP around 2006. This indicates that changes in network structure are less sensitive than simple average citation counts.

\begin{table}[h!]
\centering
\caption{Statistical summary of Cit and ASP at cluster level for WoS and Crossref datasets}
\label{tab_summary_combined}
\small
\begin{tabular}{lllllllll}
\toprule
\textbf{Dataset} & \textbf{Cluster} & \multicolumn{3}{c}{\textbf{\#Cites}} & \multicolumn{3}{c}{\textbf{ASP}} \\
\cmidrule(lr){3-5} \cmidrule(lr){6-8}
                 &                  & \textbf{Range}  & \textbf{Median} & \textbf{Mean}  & \textbf{Range}  & \textbf{Median} & \textbf{Mean}  \\
\midrule
WoS              & Medicine         & [0, 17,266]     & 1               & 6.06           & [0.5, 550.26]   & 0.53            & 0.79           \\
                 & Science          & [0, 29,906]     & 2               & 7.87           & [0.5, 2,746.82] & 0.62            & 0.87           \\
                 & Biology          & [0, 23,496]     & 3               & 8.61           & [0.5, 1,202.68] & 0.63            & 0.86           \\
                 & Engineering      & [0, 2,624]      & 1               & 4.32           & [0.5, 254.25]   & 0.52            & 0.77           \\
                 & Social Science   & [0, 11,388]     & 0               & 2.27           & [0.5, 564.27]   & 0.50             & 0.63           \\
                 & Geography        & [0, 3,339]      & 2               & 5.70           & [0.5, 171.73]   & 0.60            & 0.81           \\
                 & Arts             & [0, 1,344]      & 0               & 0.36           & [0.5, 94.24]    & 0.50             & 0.54           \\
                 & Computer Sci. & [0, 24,075]     & 1               & 3.48           & [0.5, 1,161.44] & 0.51            & 0.77           \\
                 & Psychology       & [0, 1,780]      & 1               & 4.54           & [0.5, 137.08]   & 0.54            & 0.80           \\
                 & Management       & [0, 1,173]      & 0               & 3.46           & [0.5, 100.68]   & 0.50             & 0.79           \\
                 & Law and Policy   & [0, 751]        & 0               & 1.70           & [0.5, 51.65]    & 0.50             & 0.64           \\
                 & Building         & [0, 656]        & 0               & 2.68           & [0.5, 33.28]    & 0.50             & 0.68           \\
                 & Education        & [0, 1,937]      & 0               & 2.10           & [0.5, 125.58]   & 0.50             & 0.68           \\
                 & City Dev. & [0, 739]        & 0               & 2.95           & [0.5, 48.11]    & 0.50             & 0.70           \\
\midrule
Crossref         & Medicine         & [0, 41,712]     & 1               & 5.07           & [0.5, 1,069.73] & 0.51            & 0.71           \\
                 & Science          & [0, 20,363]     & 2               & 7.22           & [0.5, 1,930.12] & 0.59            & 0.80           \\
                 & Biology          & [0, 18,343]     & 2               & 6.72           & [0.5, 1,140.10] & 0.57            & 0.75           \\
                 & Engineering      & [0, 3,599]      & 1               & 4.66           & [0.5, 170.41]   & 0.50            & 0.71           \\
                 & Social Science   & [0, 11,097]     & 0               & 1.93           & [0.5, 720.21]   & 0.50             & 0.60           \\
                 & Geography        & [0, 4,817]      & 2               & 5.87           & [0.5, 194.33]   & 0.56            & 0.75           \\
                 & Arts             & [0, 1,834]      & 0               & 0.91           & [0.5, 123.11]   & 0.50             & 0.56           \\
                 & Computer Sci. & [0, 13,226]     & 0               & 3.07           & [0.5, 645.99]   & 0.50             & 0.69           \\
                 & Psychology       & [0, 9,170]      & 1               & 4.12           & [0.5, 464.66]   & 0.52            & 0.72           \\
                 & Management       & [0, 3,636]      & 0               & 3.19           & [0.5, 141.94]   & 0.50             & 0.70           \\
                 & Law and Policy   & [0, 5,475]      & 0               & 2.67           & [0.5, 202.87]   & 0.50             & 0.62           \\
                 & Building         & [0, 759]        & 0               & 3.24           & [0.5, 35.66]    & 0.50             & 0.69           \\
                 & Education        & [0, 1,185]      & 0               & 1.40           & [0.5, 46.12]    & 0.50             & 0.59           \\
                 & City Dev. & [0, 1,192]      & 0               & 1.57           & [0.5, 70.93]    & 0.50             & 0.59           \\
\bottomrule
\end{tabular}
\end{table}

A comparison of Table \ref{tab_summary_combined} and Table \ref{tab_summary_cluster} reveals several consistent trends across all clusters after merging the WoS and Crossref datasets. First, the maximum citation values increase in all clusters, reflecting the inclusion of highly cited papers from both WoS and Crossref. For example, Medicine reaches a maximum of 46,546, and Science extends to 32,296, significantly higher than their respective pre-merge values in either dataset. This shows that the merged data reflects a significant increase in the impact of highly cited papers. Second, across all clusters, the mean citation values and ASP decline after merging compared to WoS. This reduction reflects the introduction of a larger proportion of low-citation papers, particularly from Crossref, which includes broader coverage of less-cited journals and conference proceedings. This highlights that merging results in a higher inclusion of low-citation papers. Third, the merged data shows an increased representation of low-citation articles across most clusters. This is evident from the consistent decrease or stability in median citation values, such as Medicine dropping from 1 to 0 and Arts remaining at 0, despite an expansion in the range of citations.

While the overall trends are consistent, some clusters exhibit unique behaviors after merging. First, despite the broader range of citation values and a slight decline in the mean, the Science cluster maintains its median citation value at 2, indicating a resilient core of moderately cited papers. This stands in contrast to clusters such as Medicine and Social Science, where the median either declines or remains at 0, highlighting an amplified long-tail effect in these fields. Second, the Biology cluster exhibits an interesting pattern, with the median citation value increasing slightly from 2 (Crossref) and 3 (WoS) to 3 in the merged data. This suggests that the merged dataset effectively preserves and integrates a substantial proportion of high-impact papers in this field, in contrast to the broader dilution observed in most other clusters. Third, the Arts cluster displays a uniquely low overall impact. Both the mean and median citation values remain near zero after merging, indicating that this cluster continues to be dominated by articles with minimal citation impact, regardless of the dataset.

The merging of WoS and Crossref datasets introduces significant changes across all clusters, consistently broadening citation ranges and amplifying the long-tail effect. However, specific clusters, such as Science and Biology, maintain a strong core of impactful articles, while clusters like Social Science and Arts experience a more pronounced dilution of impact due to the prevalence of low-citation papers. These findings highlight the complementary strengths of the two datasets: WoS emphasizes high-impact contributions, while Crossref provides broader coverage, creating a more diverse and inclusive merged dataset.

\subsubsection{Merged Data Completeness}

\begin{figure}[h!]
    \centering
    \includegraphics[width=\textwidth]{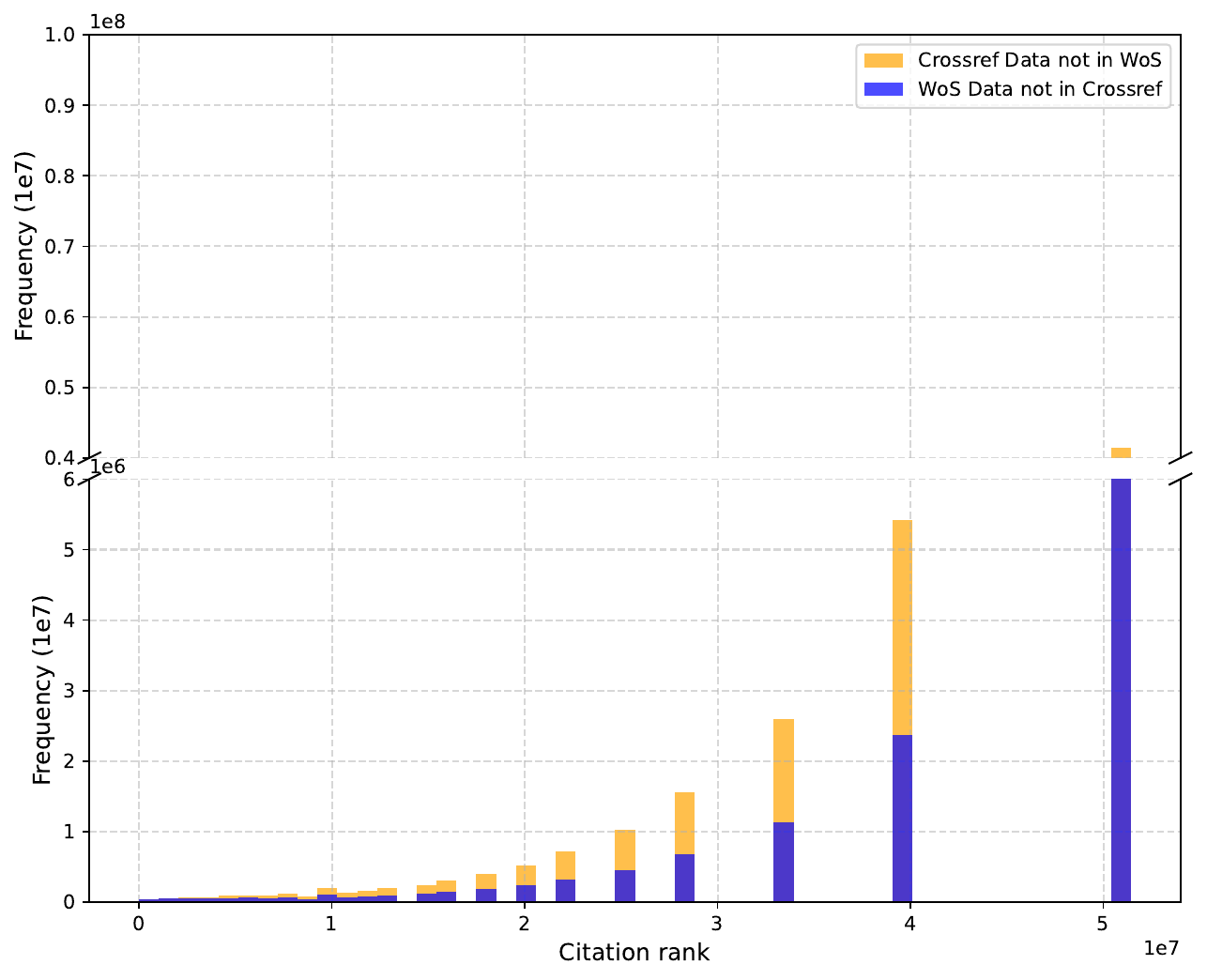}
    \caption{Citaiton rank distribution of non-overlapping papers}
    \label{Fig_Cit_distribution}
\end{figure}

First, considering the overall RCR data presented in Table \ref{tab_RCR}, the improvement in dataset citation completeness through data merging is significant. Compared to the WoS dataset, the merged data shows an average increase of 0.163 in citation data, while compared to the Crossref dataset, it demonstrates an average increase of 0.252. Secondly, data merging significantly increased the total number of papers included. For the WoS dataset, the total paper increased by 88.64\%, while for the Crossref dataset, the increase was 26.16\%. The citation rank distribution of the non-overlapping portions of the WoS and Crossref datasets within the merged data is presented in Figure \ref{Fig_Cit_distribution}.

In Figure \ref{Fig_Cit_distribution}, most of the highly ranked non-overlapping papers, with citation rankings close to zero, are sourced from WoS. As the rankings decrease, the proportion of papers from Crossref gradually rises, reflected in the increasing size of the yellow segment. This indicates that among the non-overlapping papers, WoS includes a larger proportion of highly cited papers, while Crossref provides more comprehensive coverage of low-citation papers. 

\begin{figure}[h!]
    \centering
    \includegraphics[width=\textwidth]{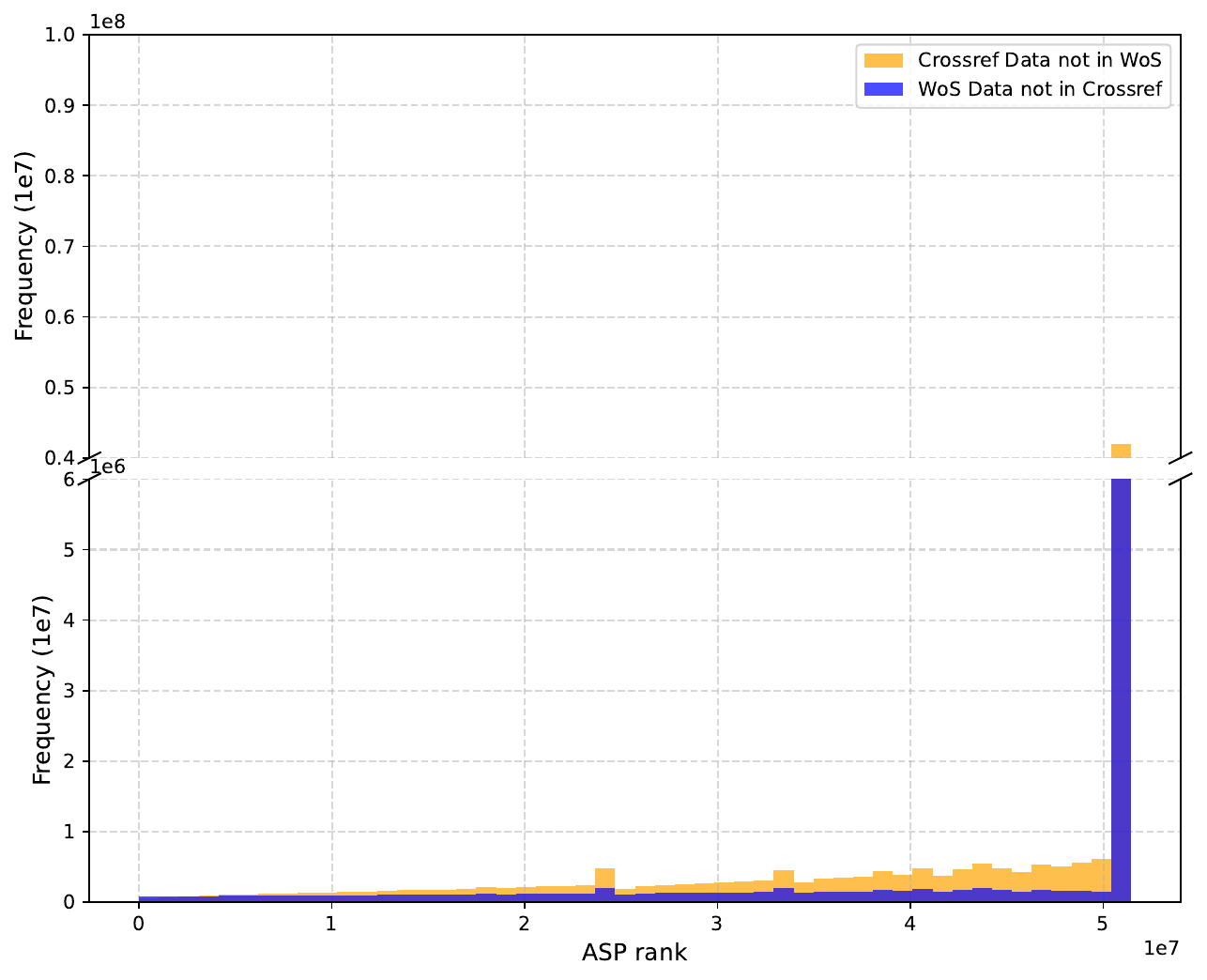}
    \caption{ASP rank distribution of non-overlapping papers}
    \label{Fig_ASP_distribution}
\end{figure}

As shown in Figure \ref{Fig_ASP_distribution}, the ASP ranking distribution is more continuous compared to the citation ranking, making it clearer that WoS has included a larger number of high-impact papers among the non-overlapping papers. Additionally, there is a slight decline in the number of low-citation papers included in WoS. The two datasets complement each other, enhancing the comprehensiveness of the data across all citation and impact levels. 

\subsubsection{Merged Data Quality}

We analyzed the ranking changes of the original WoS dataset after merging, following the method described in Section 2.3. Table 5 presents the number and proportion of the four cases. Papers with decreased impact (Case O and Case L) accounted for 69.30\% of the WoS dataset, indicating that while the merging process improved data completeness, it also inevitably introduced low-impact citations. Case G and Case D, which suggest that the newly added citations are of high quality or that the impact of existing citations has been enhanced, account for a relatively low proportion, comprising 30.70\% of the WoS dataset. 

\begin{table}[h!]
\centering
\caption{Statistical summary of four cases}
\label{tab_counts}
\small
\begin{tabular}{lrrr}
\toprule
\textbf{Case}            & \textbf{Paper Count}  & \textbf{Proportion in WoS} & \textbf{Proportion in Merged Data} \\
\midrule
G      & 16,333,732     & 26.25\%      & 13.87\%  \\
O      & 28,492,158     & 45.78\%      & 24.19\% \\
L      & 14,636,780     & 23.52\%      & 12.42\%  \\
D      & 2,770,348      & 4.45\%       & 2.35\% \\
\bottomrule
\end{tabular}
\end{table}

Among these, Case G contributes the most significantly to improving the quality of the dataset, as it further enhances the prestige of already high-impact papers by incorporating more comprehensive high-impact citations. Case D primarily contributes by enriching papers that previously relied on lower-impact citations with an increased number of high-impact citations. Therefore, the data merge contributes a significant 13.87\% improvement in data quality, top 40 papers by ASP in merged data is presented in Table \ref{tab_4Cases}.

\begin{table}[h!]
\centering
\caption{Top 40 papers by ASP in merged data}
\label{tab_4Cases}
\small
\begin{tabular}{cccccccccc}
\toprule
\textbf{$n_\text{m}$} & \textbf{$k_\text{m}$} & \textbf{$n_\text{WoS}$} & \textbf{$k_\text{WoS}$} & \textbf{$ASP_\text{m}$} & \textbf{$Cites_\text{m}$} & \textbf{Year} & \textbf{Title} & \textbf{Case} \\
\midrule
1 & 2 & 1 & 1 & 3,054.65 & 32,296 & 2008 & A Short History Of Shelx... & G \\
2 & 3 & 2 & 3 & 1,631.57 & 29,781 & 2013 & Mega6: Molecular Evolutionary... & G \\
3 & 4 & 26 & 21 & 1,405.52 & 26,916 & 2020 & Clinical Features Of Patients... & G \\
4 & 5 & 3 & 2 & 1,358.85 & 24,075 & 2016 & Deep Residual Learning For... & D \\
5 & 7 & 4 & 4 & 1,212.98 & 19,951 & 2016 & Mega7: Molecular Evolutionary... & D \\
6 & 1 & 24 & 5 & 1,146.81 & 46,546 & 2018 & Global Cancer Statistics 2018:... & G \\
7 & 8 & 42,029 & 33,716 & 1,125.47 & 17,871 & 2011 & Mega5: Molecular Evolutionary... & G \\
8 & 15 & 47 & 66 & 882.71 & 15,340 & 2020 & A Novel Coronavirus From... & G \\
9 & 11 & 11 & 9 & 873.44 & 16,582 & 2015 & Fitting Linear Mixed-Effects... & G \\
10 & 133 & 6 & 99 & 831.21 & 4,266 & 1986 & Possible High-Tc Superconductivity... & G \\
11 & 24 & 10 & 13 & 829.06 & 12,406 & 2015 & Deep Learning... & G \\
12 & 16 & 15 & 12 & 783.93 & 15,268 & 2015 & Crystal Structure Refinement With... & G \\
13 & 69 & 8 & 30 & 776.11 & 5,795 & 2015 & Going Deeper With Convolutions... & G \\
14 & 17 & 14 & 14 & 752.93 & 14,565 & 2017 & Imagenet Classification With Deep... & G \\
15 & 25 & 40 & 24,092 & 745.46 & 11,981 & 2011 & Global Cancer Statistics & G \\
16 & 30 & 5 & 15 & 703.32 & 11,339 & 2007 & Mega4: Molecular Evolutionary... & G \\
17 & 52 & 7 & 26 & 691.09 & 6,622 & 2011 & Libsvm: A Library For Support... & G \\
18 & 21 & 77 & 58 & 681.05 & 13,445 & 2020 & Clinical Characteristics Of 138... & G \\
19 & 37 & 105 & 195 & 666.13 & 8,616 & 2020 & Early Transmission Dynamics In... & G \\
20 & 26 & 116 & 84 & 639.12 & 11,916 & 2020 & Epidemiological And Clinical... & G \\
21 & 10 & 117 & 49 & 634.39 & 16,902 & 2020 & Clinical Characteristics Of... & G \\
22 & 9 & 139 & 61 & 596.17 & 16,978 & 2018 & Mega X: Molecular Evolutionary... & G \\
23 & 104 & 9 & 47 & 594.03 & 4,908 & 2003 & Single-Crystal Structure Validation... & G \\
24 & 94 & 178 & 414 & 592.19 & 5,179 & 2020 & A Familial Cluster Of Pneumonia... & G \\
25 & 12 & 17 & 6 & 577.94 & 16,300 & 2015 & Global Cancer Statistics, 2012 & L \\
26 & 88 & 13 & 44 & 564.33 & 5,296 & 2004 & Mega3: Integrated Software For... & G \\
27 & 20 & 21 & 11 & 553.50 & 13,700 & 2015 & Cancer Incidence And Mortality... & D \\
28 & 168 & 16 & 108 & 536.20 & 3,872 & 1987 & Superconductivity At 93-K In... & G \\
29 & 22 & 12 & 7 & 534.36 & 13,184 & 2011 & Global Cancer Statistics (Vol 61,... & L \\
30 & 53 & 20 & 37 & 522.75 & 6,617 & 2009 & Structure Validation In Chemical... & G \\
31 & 18 & 22 & 8 & 506.13 & 14,239 & 2016 & Cancer Statistics, 2016... & D \\
32 & 13 & 268 & 65 & 504.91 & 16,031 & 2020 & Clinical Course And Risk Factors... & G \\
33 & 23 & 29 & 18 & 498.79 & 12,923 & 2015 & Cancer Statistics, 2015... & D \\
34 & 14 & 35 & 10 & 474.82 & 15,774 & 2017 & Cancer Statistics, 2017... & G \\
35 & 83 & 19 & 36 & 456.83 & 5,399 & 2012 & Convincing Evidence From... & G \\
36 & 158 & 23 & 73 & 452.81 & 3,937 & 2014 & Dropout: A Simple Way To... & G \\
37 & 60 & 32 & 32 & 448.57 & 6,091 & 2015 & Imagenet Large Scale Visual... & D \\
38 & 32 & 60 & 24 & 437.10 & 11,173 & 2014 & Cancer Statistics, 2014... & G \\
39 & 60 & 18 & 33 & 435.52 & 6,091 & 1988 & Primer-Directed Enzymatic... & G \\
40 & 27 & 322 & 125 & 421.60 & 11,776 & 2020 & A Pneumonia Outbreak Associated... & G \\
\bottomrule
\end{tabular}
\end{table}

Since Table \ref{tab_4Cases} presents the top 40 papers based on the ASP rankings derived from the merged data, Case G constitutes the majority. Among these, the paper titled "MEGA5: Molecular Evolutionary Genetics Analysis Using Maximum Likelihood, Evolutionary Distance, and Maximum Parsimony Methods," published in 2011, exhibited the most significant improvement in ASP ranking. Its ASP increased from 11.74 in the WoS dataset, where it was ranked 42,029, to 1,125.47, securing the seventh position. The citation ranking also improved remarkably, indicating that this paper gained a substantial number of high-impact citations after the data merged. Additionally, the 2011 paper titled "Global Cancer Statistics" demonstrated a significant discrepancy between its Citation and ASP rankings in the WoS dataset, with a gap of 24,052 positions. After the data merge, this gap was reduced to just 10 positions. This result indicates that the data integration substantially improved the completeness of the paper's citation record. However, the relatively modest impact on its ASP ranking suggests that the majority of the newly added citations likely lack high influence. The best example of how data merge enhances a paper's impact is the study titled "Clinical Features of Patients Infected with 2019 Novel Coronavirus in Wuhan, China." Before the merge, this paper's ASP ranking (26) was lower than its Citation ranking (21), indicating that its citations were of relatively low impact. The merged data not only improved both rankings but also allowed the ASP ranking to surpass the Citation ranking. After the merge, its ASP and Citation rankings rose to third and fourth place. This suggests that the newly added citation records of this paper not only significantly enhanced the completeness of its citation profile but also substantially increased the influence of these citations.

Case D comprises two scenarios, both characterized by ASP rankings lagging behind Citation rankings in the WoS dataset, coupled with a decline in ASP ranking after the data merge. This indicates that in the WoS dataset, these papers were predominantly cited by lower-impact papers, and the improvement in their influence after data integration was less significant compared to other papers. However, the first scenario occurs when the ASP ranking surpasses the citation ranking post-merge, while the second scenario arises when the gap between the ASP and Citation rankings narrows. These two changes in data suggest that the influence of their citation records has slightly improved, with the improvement in the first scenario being more pronounced than in the second. For example, the 2016 paper titled "Deep Residual Learning for Image Recognition" represents Scenario 1. In the WoS dataset, its ASP ranking is 3rd, lagged behind it's second place citation ranking, after the data merge, ASP surpassed Citation, indicating an improvement in the influence of its citation record. However, the improvement was less significant compared to other papers, resulting in a decline in both ASP and Citation rankings after the merge. The 2015 paper titled "Cancer Incidence and Mortality Worldwide: Sources, Methods and Major Patterns in GLOBOCAN 2012" exemplifies Scenario 2. Its ranking gap narrowed from 10 to 7 after the data merge, indicating a slight improvement in the influence of its citation record.

There are two examples of Case L among the top 40 papers, "Global Cancer Statistics, 2012" and "Global Cancer Statistics, 2011". Both its ASP and Citation rankings declined, and the gap between the two rankings widened. This suggests that the data integration introduced low-impact citation records for this paper.

\subsection{Impact on disciplinary clusters}

\subsubsection{Data Completeness of Clusters}

Table \ref{tab_rcr_comparison} summarizes RCR across different disciplinary clusters after merging the data. Since the calculation of RCR uses the number of references in the merged data as the denominator, smaller average values of RCR\textsubscript{WoS} and RCR\textsubscript{Crossref} indicate a greater contribution of the merged data to the completeness of citations for that cluster. Therefore, for the WoS dataset, data merging contributes most significantly to the completeness of Education data, while for the Crossref dataset, it has the greatest impact on Arts. 

\begin{table}[h!]
\centering
\caption{Comparison of RCR across clusters}
\label{tab_rcr_comparison}
\small
\begin{tabular}{lrrr}
\toprule
\textbf{Cluster} & \textbf{Avg. RCR\textsubscript{WoS}} & \textbf{Avg. RCR\textsubscript{Crossref}} & \textbf{Paper Count} \\
\midrule
Education          & 0.726 & 0.759 & 326,091 \\
Arts               & 0.750 & 0.538 & 717,598 \\
Management         & 0.755 & 0.784 & 662,032 \\
Social Science     & 0.759 & 0.669 & 1,302,238 \\
City Development   & 0.772 & 0.797 & 133,455 \\
Building           & 0.784 & 0.835 & 319,746 \\
Law and Policy     & 0.784 & 0.647 & 376,769 \\
Psychology         & 0.793 & 0.728 & 741,821 \\
Computer Science   & 0.820 & 0.659 & 780,287 \\
Engineering        & 0.825 & 0.772 & 2,232,667 \\
Geography          & 0.828 & 0.769 & 2,052,484 \\
Science            & 0.832 & 0.806 & 10,585,467 \\
Biology            & 0.853 & 0.741 & 5,932,024 \\
Medicine           & 0.859 & 0.713 & 12,202,559 \\
\bottomrule
\end{tabular}
\end{table}

The data suggest that the improvement in completeness through data merging is more pronounced for smaller disciplines. For instance, disciplines such as Education, Arts, and Management exhibit significant increases in RCR values after data merging, indicating that the additional information effectively enhances citation coverage in these fields. This trend may be attributed to the relatively sparse data available for smaller disciplines in the original databases, making the supplementary effect of merging more substantial. In contrast, for larger disciplines such as Medicine and Science, which have a higher volume of papers, the impact of data merging, while still evident, results in relatively smaller increases in RCR values. This suggests that the data completeness for these fields is already relatively robust in a single dataset.

\subsubsection{Merging Performance of Clusters}

Table \ref{tab_cluster_cases} presents the proportion of data for the four types of cases within each cluster. Case G demonstrates the impact of data fusion on data quality improvement. The analysis reveals that Geography (36.16\%, 1,005,288 papers) shows the most significant proportional improvement, indicating that this cluster benefits the most from data merging in terms of data quality enhancement. Management (35.68\%, 379,591) and Science (33.13\%, 4,599,563) follow closely, with Science exhibiting the highest absolute frequency, underscoring its substantial contribution to overall data quality improvement across a large dataset. These results suggest that Geography, Science, and Management are key cluster where data merging yields the most pronounced benefits. Although Medicine (23.04\%, 4,374,294) ranks second only to Science in the absolute number of papers under Case G, it has the highest proportion of papers in Case O (46.03\%, 8,739,648). This indicates that while Medicine contributes to the quality improvement of the merged data in terms of quantity, it also introduces the largest number of low-impact papers and citations. This phenomenon primarily arises because Medicine has the largest data base (18,986,394 papers), exceeding the second-largest cluster, Science, by 36.77\%.

\begin{table}[h!]
\centering
\caption{Statistical summary of four cases across clusters}
\vspace{2em} 
\label{tab_cluster_cases}
\small
\adjustbox{trim=0in 0in 0in 0in, clip, margin*=-1in}{
\begin{tabular}{lrrrrr}
\toprule
\textbf{Cluster} & \textbf{G} Prop. (Freq.) & \textbf{O} Prop. (Freq.) & \textbf{L} Prop. (Freq.) & \textbf{D} Prop. (Freq.) & \textbf{Total} \\
\midrule
Geography          & 36.16\% (1,005,288)  & 34.50\% (959,325)   & 22.75\% (632,653)   & 6.58\% (183,082)   & 2,780,348 \\
Management         & 35.68\% (379,991)    & 52.74\% (561,669)   & 9.52\% (101,337)    & 2.06\% (21,909)    & 1,064,906 \\
Science            & 33.13\% (4,599,563)  & 32.79\% (4,551,689) & 28.42\% (3,945,866) & 5.66\% (785,223)   & 13,882,341 \\
City Dev.   & 31.91\% (73,736)     & 52.80\% (122,028)   & 12.15\% (28,087)    & 3.14\% (7,256)     & 231,107 \\
Psychology         & 30.96\% (343,921)    & 48.72\% (541,210)   & 15.74\% (174,867)   & 4.59\% (50,950)    & 1,110,948 \\
Computer Sci.   & 29.90\% (802,642)    & 49.71\% (1,334,616) & 17.28\% (463,984)   & 3.11\% (83,442)    & 2,684,684 \\
Education          & 28.52\% (166,048)    & 59.11\% (344,123)   & 10.39\% (60,509)    & 1.97\% (11,455)    & 582,135 \\
Engineering        & 28.41\% (1,612,764)  & 52.26\% (2,966,863) & 15.81\% (897,519)   & 3.52\% (199,800)   & 5,676,946 \\
Building           & 27.60\% (181,519)    & 61.56\% (404,871)   & 8.28\% (54,487)     & 2.55\% (16,795)    & 657,672 \\
Biology            & 25.10\% (2,031,905)  & 32.03\% (2,592,834) & 36.63\% (2,964,766) & 6.24\% (505,300)   & 8,094,805 \\
Medicine           & 23.04\% (4,374,294)  & 46.03\% (8,739,648) & 26.43\% (5,018,872) & 4.50\% (853,580)   & 18,986,394 \\
Law and Policy     & 21.18\% (173,473)    & 68.48\% (560,750)   & 8.72\% (71,384)     & 1.62\% (13,267)    & 818,874 \\
Social Science     & 12.82\% (377,785)    & 80.84\% (2,382,435) & 5.35\% (157,540)    & 0.99\% (29,233)    & 2,946,993 \\
Arts               & 7.41\% (195,867)     & 90.21\% (2,385,997) & 2.09\% (55,293)     & 0.29\% (7,659)     & 2,644,816 \\
\bottomrule
\end{tabular}
}
\vspace{6em}
\end{table}

Case O represents data merging results in a decline in data quality due to the incorporation of low-impact papers and citations. Arts (90.21\%, 2,385,997 papers), Law and Policy (80.42\%, 560,750), and Social Science (80.84\%, 2,494,339) show the most significant proportional impact, suggesting that these clusters are particularly vulnerable to low-quality data. Case L represents a more severe decline in data quality compared to Case O. The most affected cluster is Biology, with 36.63\% (2,964,766 papers) of its data falling into this category, which is also the highest proportion within the cluster. This indicates that after data merging, the most significant declines in data quality occurred in the Biology, Arts, and Social Science.

\begin{table}[h!]
\centering
\caption{Top 20 papers by ASP in merged Science cluster}
\vspace{5em} 
\label{tab_Science_ASP}
\small
\adjustbox{trim=0in 0in 0in 0in, clip, margin*=-1in}{
\begin{tabular}{cccccccccc}
\toprule
\textbf{$n_\text{m}$} & \textbf{$k_\text{m}$} & \textbf{$n_\text{WoS}$} & \textbf{$k_\text{WoS}$} & \textbf{$ASP_\text{m}$} & \textbf{$Cites_\text{m}$} & \textbf{Year} & \textbf{Title} & \textbf{Case} \\
\midrule
1 & 2 & 1 & 1 & 3,054.65 & 32,296 & 2008 & A Short History Of Shelx ... & G \\
10 & 133 & 6 & 99 & 831.22 & 4,266 & 1986 & Possible High-Tc Superconductivity ... & G \\
12 & 16 & 15 & 12 & 783.93 & 15,268 & 2015 & Crystal Structure Refinement With ... & G \\
23 & 104 & 9 & 47 & 594.04 & 4,908 & 2003 & Single-Crystal Structure Validation ... & G \\
28 & 168 & 16 & 108 & 536.20 & 3,872 & 1987 & Superconductivity At 93-K In ... & G \\
39 & 60 & 18 & 33 & 435.52 & 6,091 & 1988 & Primer-Directed Enzymatic ... & G \\
45 & 130 & 25 & 70 & 385.69 & 4,301 & 2010 & Review Of Particle Physics ... & G \\
48 & 115 & 30 & 57 & 355.25 & 4,605 & 2007 & The Rise Of Graphene ... & G \\
49 & 173 & 28 & 95 & 342.70 & 3,848 & 2004 & Review Of Particle Physics ... & G \\
62 & 393 & 37 & 193 & 313.14 & 2,615 & 2002 & The Cambridge Structural Database ... & G \\
80 & 217 & 566 & 674 & 277.21 & 3,471 & 2012 & Wingx And Ortep For Windows ... & G \\
82 & 85 & 49 & 43 & 270.67 & 5,344 & 2012 & Review Of Particle Physics ... & D \\
84 & 314 & 46 & 146 & 265.75 & 2,927 & 2000 & Review Of Particle Physics ... & G \\
92 & 671,300 & 68 & 524,397 & 249.63 & 74 & 2012 & Estimating Divergence Times In ... & G \\
94 & 75 & 69 & 39 & 248.08 & 5,545 & 2014 & Review Of Particle Physics ... & D \\
96 & 65 & 2,455 & 1,081 & 246.26 & 5,925 & 2015 & Sheltx - Integrated Space-Group ... & G \\
98 & 2,212 & 59 & 1,262 & 244.95 & 1,125 & 2004 & Electric Field Effect In ... & G \\
104 & 232 & 63 & 113 & 238.32 & 3,358 & 2008 & Iron-Based Layered Superconductor ... & G \\
108 & 131 & 65 & 63 & 236.33 & 4,290 & 2008 & Review Of Particle Physics ... & D \\
109 & 2,151 & 91 & 1,613 & 236.22 & 1,140 & 1987 & Evidence For Superconductivity ... & G \\
\hline
\end{tabular}
}
\vspace{6em}
\end{table}

Table \ref{tab_Science_ASP} selected Science, which has both a high proportion and a large absolute number of papers in Case G, to analyze the impact of top papers in this cluster. The ASP ranking of the top 20 papers in Science demonstrates an improvement in data quality after merging the data (All Case G and Case D). Among these papers, the one with the most significant improvement in ASP ranking is the 2015 publication, "Sheltx - Integrated Space-Group," which advanced from a ranking of 2,455 to 96, indicating a substantial enhancement in the article's prestige. The paper with the largest ASP and Citation ranking disparity is the 2012 publication, "Estimating Divergence Times in Large Molecular Phylogenies." Even prior to data merging, the ranking disparity was 524,329, suggesting that it already exhibited a pattern of limited but high-impact citations within the WoS dataset. After merging, this disparity increased to 671,208, reflecting a further amplification of its citation list's influence.

While data merging generally has a positive impact on high-impact papers in Science, most of these papers experience a decline in both ASP and citation rankings. The positive effect is more evident in the increased gap between citation and ASP rankings. This suggests that the new citations added by data merging are relatively few, leading to a general decline in citation rankings. Evidence for this is provided in Table \ref{tab_rcr_comparison}, where Science ranks third in average RCR within the WoS dataset, indicating a high-impact but limited number of new citations. These high-impact citations help soften the decline in ASP rankings, resulting in a greater disparity between the two rankings after merging.

\section{DISCUSSION}

Our study focuses on comparing the differences between scientific datasets from different sources and examining the impact of data merging. We evaluate the quality of datasets primarily from two perspectives: Reference Coverage Rate (RCR), which assesses the completeness of references included in the dataset, and Article’s Scientific Prestige (ASP), which measures the impact of the papers within the dataset. Additionally, we analyze changes in ASP and citation rankings to categorize the impact of data merging into four cases. Furthermore, we examine the effects of merged data from both the individual paper and cluster perspectives.

\subsection{Contributions}

First, we observe significant differences in data quality between the datasets, even when covering the same article. The WoS dataset outperforms Crossref in terms of reference coverage rate (RCR) and article scientific prestige (ASP), particularly excelling in its coverage of high-impact literature. However, the Crossref dataset provides complementary value by offering broader coverage of lower-impact literature, thereby addressing the limitations of the WoS dataset in terms of data completeness. 

Second, we find that data merging significantly enhances citation completeness. Through data integration, the overall reference coverage rate improves markedly, particularly in smaller cluster such as Education, Arts, and Management, where the merged data effectively addresses the citation coverage gaps of the original datasets. 

Third, we observe that data merging leads to greater stratification in citation quality. The integration process introduces a large number of new citations, which partially enhance the scientific prestige of high-impact literature. However, it also inevitably incorporates low-impact citation data, resulting in a polarization of overall data quality. This phenomenon is particularly pronounced in disciplines with large datasets, such as Medicine and Biology.

\subsection{Limitations}

The ideal calculation of RCR would use the actual number of references cited by a paper as the denominator and the number of references included in a single dataset as the numerator, providing an accurate coverage ratio. However, since our datasets do not provide the actual number of references for each paper, we used the closest available approximation: the number of references in the merged dataset as the denominator for the calculation. Although this method allows us to compare the relative completeness of references among WoS, Crossref, and merged data, the results do not accurately reflect the absolute completeness. We can only determine the contribution of merged data to improving completeness; whether the merged data achieves full completeness remains further investigation.

This study focused exclusively on two data sources, WoS and Crossref, while more than 30 commonly used data sources exist in science of science research. Future studies could expand to include additional sources, such as Scopus, Dimensions, and Microsoft Academic Graph (MAG), to further validate the generalizability of data merging for quality enhancement.

\section{CONCLUSION}

This study analyzed the impact of data quality on scientometric research by merging WoS and Crossref datasets. The findings indicate significant differences in data quality between the two datasets: WoS demonstrates superior coverage of high-impact literature and article scientific prestige (ASP), while Crossref provides complementary coverage of low-impact literature. Data merging was found to be highly effective in improving overall citation coverage, particularly in smaller disciplines, and positively influenced the scientific prestige of high-impact literature, enhancing the completeness and stability of the citation network. However, the impact of data merging varies across cluster. Fields such as Science, Biology, and Medicine benefit the most, while Social Sciences, Arts, and Law are more susceptible to the inclusion of low-impact literature. Furthermore, while data merging addresses citation coverage gaps, it also introduces a portion of low-impact citations, leading to a dual effect on data quality. Future research must focus on achieving a balance between enhancing data quality and filtering low-quality data to optimize the benefits of data integration.

\section*{Author Contributions}

Guoyang Rong: Data curation, Methodology, Formal analysis, Visualization, Writing—original draft,  Writing—review \& editing. Ying Chen: Conceptualization, Methodology, Project administration, Resources, Supervision, Writing – review \& editing.  Thorsten Koch: Data curation, Resources, Software, Funding acquisition. Keisuke Honda: Data curation, Resources. \vspace{\baselineskip}

\section*{Competing Interests}

The authors have no competing interests.

\section*{Funding Information}

Part of the work has been co-funded by the European Union (European Regional Development Fund ERDF, fund number: STIIV-001 ). Part of the work has been conducted in the Research Campus MODAL funded by the German Federal Ministry of Education and Research (BMBF) (fund numbers 05M14ZAM, 05M20ZBM).

\section*{Data Availability}
The data supporting the findings of this study were retrieved from the Web of Science and Crossref, accessed through the Institute of Statistical Mathematics and Zuse Institute Berlin subscriptions.

\printbibliography

\end{document}